# The impact of Use Cases in real-world software development projects: A systematic mapping study.

José L. Barros-Justo[a, 1], Fabiane B.V. Benitti[b], Saurabh Tiwari[c]

[a] *School of Informatics (E.S.E.I.), University of Vigo, (32004) Ourense, Spain*
[b] *Universidade Federal de Santa Catarina (UFSC), Florianópolis, Santa Catarina, Brazil*
[c] *Dhirubhai Ambani Institute of Information and Communication Technology (DA-IICT), Gandhinagar, India*

ABSTRACT

**Context**: There is abundant literature on the application of UML Use Cases. However, the impact that these applications have had on real projects (in industry) is not well known. It is necessary to know what the impact of the Use Cases really is, so that both, researchers and professionals can make the most pertinent decisions.

**Objective**: To identify and classify the positive and negative impacts of using Use Cases in real-world settings.

**Method**: We conducted a systematic mapping study. The search strategies retrieved a set of 4,431 papers out of which 47 were selected as primary studies. We defined four facets to classify these studies: 1) the positive impact (advantages), 2) the negative impact (disadvantages), 3) the industry's domain and 4) the type of research reported.

**Results**: Our study identified eight categories of advantages related to the application of Use Cases. The most mentioned were estimation, analysis and automation. These advantages had a heterogeneous distribution along the years. On the other hand, the granularity of the scenarios described in the Use Cases, the lack of a standardized format for specifying requirements, and the lack of appropriate guidelines for analysing them were the most mentioned disadvantages. We identified a variety of industry domains, grouped into seven categories. As we can expect most of the papers report experiences in the Information Technology domain, followed by financials applications. Almost half the papers applied evaluation research, including an empirical validation. Only one third of the analysed papers reported threats to validity, the most mentioned being generalizability (38%).

**Conclusions**: Use Cases have proven to be a useful tool in software development, particularly during the early stages. The positive effects far outweigh the few negative effects reported, although this may be due to the researchers' tendency of not reporting negative results.

## 1. Introduction

Use Cases are defined as a mechanism to elicit, to specify and to validate software requirements [1]. A Use Case describe the requirements of a system in terms of the main actors (external elements that interact with the system) and their goals [2]. Traditionally, Use Cases have been used to represent the functionality of a system, but they can also help to elicit requirements such as Reliability, Safety, Security, and Usability (Non Functional Requirements) [1]. The granularity of Use Cases, in terms of their structure, has made them useful not only during the requirements elicitation/validation process, but also in several software development life-cycle activities, such as planning and estimation [3–5], analysis model generation [2,6–9], and test case generation [10–12].

---

[1] Corresponding author. Tel.: +34 988 387 029
Email addresses: jbarros@uvigo.es (José L. Barros-Justo), fabiane.benitti@ufsc.br (Fabiane B.V. Benitti), saurabh_t@daiict.ac.in (Saurabh Tiwari).

While the application of Use Cases in software engineering is well documented by abundant reports in journals and conferences [13–17], there are few systematic studies reporting their benefits and the impact they had in industry settings. For practitioners, it is important to have information regarding the effect that Use Cases will have on their projects, for example their impact on cost and development time. This information will allow them to make better decisions and achieve management support more easily. On the other hand, researchers need to validate their theoretical proposals in real projects. The result of the adoption of techniques and tools in the industry provides important information to decide on future lines of research.

In recent years, the paradigm of Evidence-Based Software Engineering (EBSE) and its two main research methods, the systematic mapping studies (SMS) and the systematic literature reviews (SLR) have gained great notoriety [18–21]. Both tools provide a research method that systematically identifies, classifies and analyses the data reported in scientific publications (the evidence). The main goal of a SMS is to identify and classify the existent data [22,23], while a SLR aims to synthesize new information based on the extracted data [24–27].

This work presents a SMS aimed at identifying and classifying the impact of using Use Cases in real-world software development projects. The contributions of this study are:

- C1: Identify the impact on software development due to the application of Use Cases technology,
- C2: Classify the impacts according to their outcome (advantage or disadvantage) and the context were they have occurred (industry's domain),
- C3: Identify research gaps in the existing literature and therefore, helps to plan for future research,
- C4: Highlight some demographic data, such as active researchers, organizations and countries, and most frequent publication venues.

The rest of the study is structured as follows: section 2 presents the related work and briefly analyse the conclusions of other secondary studies; section 3 depicts the research method and details the protocol that was followed to guide the main activities of this study; section 4 shows, analyses and discusses the results. Section 5 review the threats to validity and, finally, section 6 presents the conclusions and proposals for future work.

## 2. Related work

We ran search queries to find relevant related work, i.e. secondary studies focused in the practical application of Use Cases. These queries were executed in ACM DL, IEEE Xplore, SCOPUS and Web of Science, using the following search string:

*(Use Case) AND (impact\* OR benefit OR \*advantage OR \*creas\* OR improve\* OR \*crement\*) AND (industr\* OR business OR organization OR firm OR compan\*) AND (review OR mapping OR tertiary)*

These searches retrieved 52 studies from which we selected three works [10,28,29] that fulfilled our inclusion criteria:

- Secondary or tertiary studies,
- Focused on the application of Use Cases.

Using the previous selected works as seeds, we performed backward and forward snowballing to identify and analyse 262 more studies. Only one new study was included [30]. Table 1 summarize the key information found in the following four secondary studies.

Santos, Andrade and Santos Neto [28] presenting a systematic mapping study about Software Product Lines (SPL) variability description in textual Use Cases. The study started with 1,892 primary studies published until December 2013, but only nine studies were finally considered as relevant. They formulated three research questions: (i) Which are the templates for the textual Use Case of an SPL? (ii) How could SPL variability be modelled in textual Use Cases? and (iii) Which variability types can be modelled in textual Use Cases of an SPL? The authors found nine Use Case templates and four different ways to describe SPL variabilities in a Use Case description. From these templates, only one dealt with all the five variability types identified. The authors did not find any experimental study comparing these templates in terms of ease of use or comprehensibility, but, based on the evidence identified, they conclude that the specification of the SPL functional requirements with Use Case could offer enough details to help other development activities like design and testing.

Tiwari and Gupta [29] conducted a systematic literature review to report the advancements about the Use Case specification research. The study analysed 119 primary papers published between 1992 and February 2014. Four research questions were formulated in order to identify: (i) the evolution of the Use Cases, (ii) their applications, (iii) quality assessments and (iv) open issues and future directions. The results of the study revealed that twenty different instances of Use Cases have been proposed in different contexts and for different purposes. The authors conclude that the applicability of Use Cases in a specific context needs to be investigated in more detail. The authors also reported several open issues, challenges and future directions, specifically, they highlighted the need to analyse the industrial relevance of the Use Cases.

In 2016, Qazi, Rauf and Minhas [10] conducted a systematic literature review of Use Cases based software testing techniques. Their goal was to synthesize the approaches (RQ1) to Use Cases based software testing by means of several parameters (RQ2). The study analysed 45 primary papers from 2001 to 2010. The approaches (RQ1) were categorized according to their strengths (proposed steps, tool support and level of testing) and weaknesses (complexity in terms of cost, time and training). The analysed parameters (RQ2) included level of testing, used algorithm, tool support, level of automation, software domain and complexity metrics, among others. The authors conclude that Use Cases can help to conduct requirements based testing, system testing at high-level design (software architecture) and low-level design (integration testing).

El-Attar and Miller [30] conducted a systematic literature review to identify and combine guidelines, suggestions and techniques, provided by researchers and practitioners, to construct high quality Use Case models. They gathered information about: a) UC modelling, its notation, syntactical rules and semantics; b) how to apply UC modelling, best practices, quality attributes and patterns; and c) pitfalls, mistakes and poor quality attributes. The study focused in five quality attributes of a UC model: consistency, correctness and completeness, fault-free, analytical and understandability. Eighty-four primary papers were selected and analysed to produce a final list of 61 unique rules (heuristics), which were finally compiled into a set of 21 anti-patterns (what not to do in UC modelling).

Table 1 shows a summary of key issues extracted from the related works analysed in this section. Our work differs from these studies in the following aspects:

- It focuses on real-world industrial environments, where there is a clear lack of empirical data.
- All the reported impacts (positive and negative) were included (open list), and
- The period considered is up to September 30, 2018.

*Table 1   Summary of relevant data extracted from the secondary studies*

| Reference | Goal | Conclusions | Period | #Papers |
|---|---|---|---|---|
| [28] | offers a summary of existing Use Case templates as a result of a SMS with a focus on the description of variations in textual Use Cases. | Nine Use Case templates and four different ways to describe SPL variabilities in a Use Case description. | Up to 12/2013 | 9 |
| [29] | examine the existing literature for the evolution of the Use Cases, their applications, quality assessments, open issues, and the future directions | Use Cases have been evolved from initial plain, semi-formal textual descriptions to a more formal template. The authors have also highlighted various issues and challenges related to Use Case specifications. | From 1992 to 02/2014 | 119 |
| [10] | Identify approaches to Use Cases based software testing. | Use Cases can be helpful at any phase of software development and for any type of testing strategy. | From 2001 to 2010 | 45 |
| [30] | Guidelines and techniques to produce high quality UC models. | 21 anti-patterns reflecting what not to do in UC modelling, | Up to 2009 | 84 |

## 3. Research method

In recent years, the paradigm of Evidence-Based Software Engineering (EBSE) and its two main research methods, the systematic mapping studies (SMS) and the systematic literature reviews (SLR) have gained great notoriety [18–21]. Both tools provide a research method that systematically identifies, classifies and analyses the data reported in scientific publications (the evidence). The main goal of a SMS is to identify and classify the existent data [22,23], while a SLR aims to synthesize new information based on the extracted data [24–27].

We followed the guidelines in [19,23] to plan and conduct this SMS. Recommendations for the reporting were taken from [31]. The next sections present the general goal and research questions and details the steps of the conducting phase.

### 3.1. Specify goal and research questions

The goal of this mapping study was established as:

*Identify and classify the impacts of using Use Cases in real-world software development projects*

From this goal, we derived a set of Research Questions (RQs) and Demographic Questions (DQs) as suggested in [32,33].

**Research questions**

- *RQ1: Which are the most frequently reported advantages or disadvantages of using Use Cases in software development in real-world settings? And, How have they evolved over time?*
- *RQ2: Which are the most reported industry's domains?*
- *RQ3: Which research types were applied?*
- *RQ4: Were threats to validity reported?*

**Rationale:** In 1986, Ivar Jacobson formulated the concept of Use Cases for the first time, and presented the techniques for its application in software development [34]. Therefore, the industry has been in possession of this powerful technique for more than 30 years. However, it is still unknown what the advantages and the disadvantages (RQ1) are associated with the application of the Use Cases in real software development projects [35], that is, in industrial environments (RQ2) [36]. The answers to these questions will help practitioners make sound, evidence-based decisions to improve the performance of their development processes and the quality of the final products.

Finally, as researchers, we need to know the quality of the data we are using. For this purpose, the questions RQ3 and RQ4 allow investigating in the formal scientific aspects of the primary works, analysing both the research method used and the potential threats to the validity.

**Demographic questions**

- *DQ1: Who are the most active researchers?*
- *DQ2: Which are the most active organizations?*
- *DQ3: Which are the most active countries?*
- *DQ4: Which are the Top publication venues?*

**Rationale:** Demographics typically represent characteristics of a population (population here refers to the domain information about Use Case research). The rationale behind each of the DQs includes characteristics such as active researchers, active organisations (clustered by their type: academy or industry), active countries, and top publication venues (including journals and conferences). This set of DQs is frequently suggested in several guidelines related to the preparation of secondary studies [27,33,37,38].

To help us classify the different research types reported in the primary studies, we have developed a list of conditions and the values that must be met to select the proper research type. Table 2 shows these conditions and help to decide whether a primary paper reports on one of the research types proposed by Wieringa et al. [39]. Notice, for example, that the only difference between Evaluation and Validation research is the value of the first condition (first row: "Used in practice"). Since our study focuses on the application of Use Cases in real world projects, the condition "Used in practice" (first row in Table 2) should be true (T) or irrelevant (*). Therefore, the possible values for type of research in our set of selected studies are Evaluation research, Solution proposal or Experience report (grey coloured cells in Table 2).

*Table 2   Research type classification (T = True, F = False, *= irrelevant)*

| Evaluation | Validation | Philosophical | Solution | Opinion | Experience |
|---|---|---|---|---|---|

|  | research | research | paper | proposal | paper | report |
|---|---|---|---|---|---|---|
| **Conditions** | | | | | | |
| Used in practice | T | F | F | * | F | T |
| Novel solution | * | * | F | T | F | F |
| Empirical evaluation | T | T | F | F | F | F |
| Conceptual framework | * | * | T | * | F | * |
| Author's opinion | F | F | F | F | T | F |
| Author's experience | * | * | F | * | F | T |

The meaning of the "conditions" (first column in Table 2) is as follows [39]:

- "Used in practice": This is the investigation of a problem in RE practice or an implementation of an RE technique in practice (not in a lab or in an academic context).
- "Novel solution": The technique must be novel, or at least a significant improvement of an existing technique.
- "Empirical evaluation": Causal properties are studied empirically, such as by case study, field study, field experiment, survey, etc. Logical properties are studied by conceptual means, such as by mathematics or logic.
- "Conceptual framework": These Papers sketch a new way of looking at things.
- "Author's opinion": report the author's opinion about what is wrong or good about something.
- "Author's experience": In these papers, the emphasis is on what and not on why. The experience may concern one or more projects, but it must be the author's personal experience. The paper should contain a list of lessons learned. Papers in this category will often come from industry practitioners or from researchers who have tried their tools or methods in practice.

### 3.2. Search for primary studies

The search consists of two different strategies: *Automated search* ran in publishers and indexers databases and *backward and forward* snowballing using the studies selected by the automated search as seeds. Figure 1 summarizes the search process and shows the results obtained from each strategy.

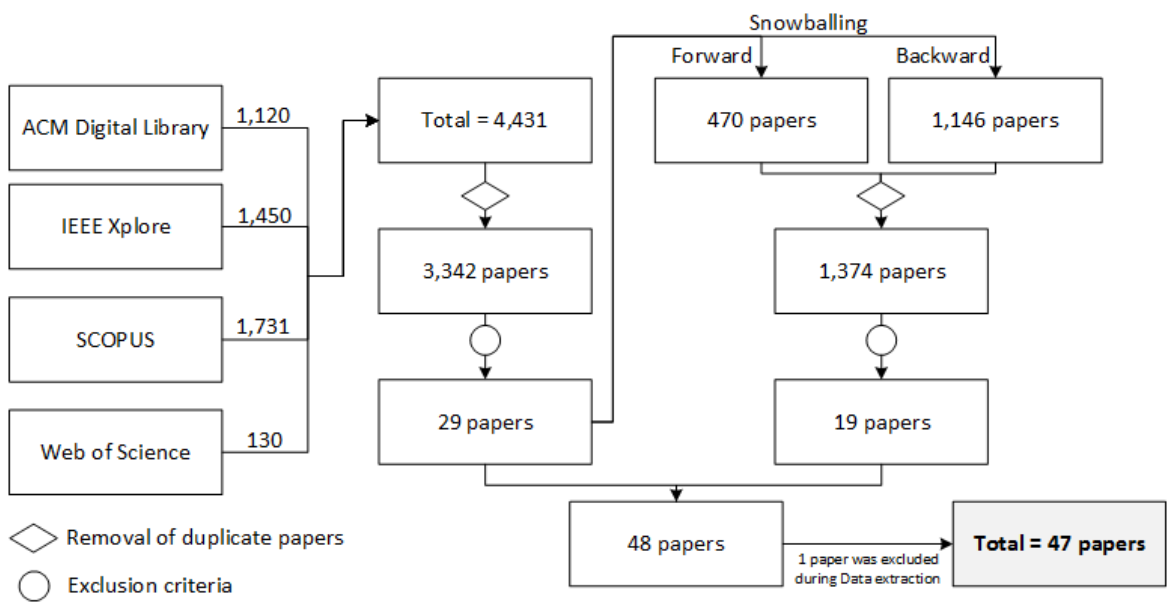

*Figure 1   Results from the search strategies*

**Automated search**

To conduct the automated search, a search string was built. Special care was taken in this step, since it has been reported that systematic reviews fail because of an inadequate search string construction [40–44].

To create the search string, we first derived the major search terms from the main goal and the set of RQs and DQs. After that, we conducted a pilot search in SCOPUS, using these major terms to identify other relevant keywords, synonyms and alternative spellings. Finally, we applied the PICOC method (Population, Intervention, Comparison, Outcome and Context), as suggested by Kitchenham et al. [19] and connect the resulting terms using Boolean operators. The extracted key terms were:

- Goal: *impact, Use Case.*
- RQ1: *benefit, advantage, increase, improve, increment, disadvantage, decrease, decrement.*
- RQ2: *industry, firm, business, organization, company*.
- DQ4: *journal, conference, workshop*.

The goal of our study is to assess the impact of Use Cases (both, advantages and disadvantages) in real-world scenarios. RQ1 directly address this goal, while RQ2 is aimed to describe the context. RQ3 and RQ4 are aimed to evaluate the quality of the retrieved evidence (do not add to search key terms). Similarly, for DQs, answers can only be found after the identification of primary studies (no need to derive search key terms for them), except for DQ4, which has been used as a filter (in the search) and as a classification facet.

PICOC distribution of key terms and extra terms to limit the Context facet:

- Population: *Use Cases.*
- Outcomes: *impact, benefit, advantage, increase, improve, increment, disadvantage, decrease, decrement.*
- Context: *industry, firm, business, organization, company.*
- Limits: *journal article OR conference paper; subject area: computer science; Language: English.*

After reviewing the search results for different combinations of key terms, the final search string was established as:

(Use Case) AND (impact* OR benefit OR *advantage OR *creas* OR improve* OR *crement*) AND (industr* OR business OR organization OR firm OR compan*) AND (Journal article OR Conference paper) AND (SUBJAREA: Computer Science) AND (LANGUAGE : English)

To avoid bias we included four complementary electronic data sources (EDS) which are well known among researchers and academics: ACM Digital Library (ACM DL), IEEE Xplore, SCOPUS and Web of Science, as suggested in [45]. The first two covered the most important journals and conferences in the field of software engineering [41], while the last two are recognized as the largest general indexing services, including papers published by ACM, IEEE, Elsevier, Springer and Wiley [46].

Finally, we tailored the search string. Table 3 shows the final syntaxes for each EDS, given their particular query language and searching restrictions.

Table 3   Search strings tailored to each EDS.

| EDS | Search string |
|---|---|
| ACM DL | query: { ( "Use Case" ) AND ( impact* OR benefit OR *advantage OR *creas* OR improve* OR *crement* ) AND ( industr* OR business OR organization OR firm OR compan* ) } |
| IEEE Xplore | ("Abstract":"Use Case") AND ("Abstract":impact OR "Abstract":benefit OR "Abstract":*advantage OR "Abstract":*creas* OR "Abstract":improve*) |
| SCOPUS | TITLE-ABS-KEY ( ( "Use Case" ) AND ( impact* OR benefit OR *advantage OR *creas* OR improve* OR *crement* ) AND ( industr* OR business OR organization OR firm OR compan* ) ) AND ( LIMIT-TO ( DOCTYPE , "cp " ) OR LIMIT-TO ( DOCTYPE , " ar " ) ) AND ( LIMIT-TO ( SUBJAREA , "COMP " ) ) AND ( LIMIT-TO ( LANGUAGE , "English " ) ) |
| Web of Science | TS=(("Use Case") AND ( impact* OR benefit OR *advantage OR *creas* OR improve* OR *crement* ) AND ( industr* OR business OR organization OR firm OR compan* )). Refined by: WEB OF SCIENCE CATEGORIES: ( COMPUTER SCIENCE SOFTWARE ENGINEERING ) |

Table 4 shows the outcomes from the automated search strategy.

Table 4   Automated search results

| EDS | Retrieved | Unique |
|---|---|---|
| ACM DL | 1,120 | 427 |
| IEEE Xplore | 1,450 | 1,177 |
| SCOPUS | 1,731 | 1,718 |
| Web of Science | 130 | 20 |
| Total = | 4,431 | 3,342 |

**Snowballing search**

Snowballing refers to the process of using the reference list of a paper (backward snowballing) or the citations to this paper (forward snowballing) to identify additional papers. We conducted a snowballing search using the papers selected from the automated search strategy as the initial set of seeds, following the guidelines proposed in [47].

The backward strategy retrieved 1,146 works, while additional 470 papers came from the forward process. We removed 242 duplicates. Three iterations were needed to complete the snowballing process. The final output was 19 works added to the set of previously selected papers. Table 5 presents the summary of the outputs from this search strategy.

Table 5   Snowballing search results

| Iteration | References | Citations | Duplicates | Selected as Seeds |
|---|---|---|---|---|
| First | 683 | 340 | 187 | 13 |
| Second | 286 | 100 | 35 | 6 |
| Third | 177 | 30 | 20 | 0 |
| Total works = | 1,146 | 470 | 242 | 19 |

## 3.3. Selection of primary studies

We started by eliminating duplicates in the set of retrieved papers. This preliminary task ensures that we do not analyse the same paper more than once. After that, all papers were gathered into a single spreadsheet and a two-step process was performed:

Step 1: the results were screened on title, keywords, and metadata only, to exclude papers filling any of the following EC (Exclusion Criterion):

- EC1: Not written in English,
- EC2: Short paper (less than 4 pages length)
- EC3: Not peer-reviewed publication,
- EC4: Not a primary study (secondary/tertiary studies were analysed in the related work section),
- EC5: Books, editorials, tutorials, panels, poster sessions, prefaces, opinions, letters, slides, key notes, web pages, forewords and any other work that can be considered as grey literature,
- EC6: PhD or Master Theses, under the assumption that relevant publications, resulting from the research covered by these studies, were already published as research papers on peer-reviewed venues.
- EC7: for studies having conference and journal versions, we excluded the oldest one (normally the Conference version); if they have the same date, then we excluded the less detailed one.

Step 2: Papers not excluded by Step 1 were evaluated based on their abstracts and, if necessary on a full-text reading, to exclude those that were not within the scope, i.e., irrelevant papers that were retrieved due to poor execution of the search strategies. This scope criterion was subdivided into three complementary criteria:

- o EC_Pop: Not about the application of Use Cases [Not in Population],
- o EC_Out: Not about any impact (positive or negative) [Not in Outcome],
- o EC_Con: Not an industrial setting [Not in Context]

The application of the exclusion criteria was done by:

- One reviewer, to conduct the removal of duplicates (first author)
- A pair of reviewers, to conduct Step 1 (first and second author)
- Two pairs of reviewers, to conduct Step 2: one pair (second and third author) analysed the first half of the selected papers from Step 1 and the second pair (first and third author) the other half. On a second round, the pairs of reviewers interchange their assigned halves.

To deal with disagreements, we applied the inclusive criteria proposed in [23]. For example, if one Reviewer is uncertain about excluding a paper and the other Reviewer consider that the paper should be included (category "B"), the paper is included. The only categories that implies the exclusion of a paper are "E" (one of the Reviewers vote for Exclusion and the other Reviewer is Uncertain) and "F" (both Reviewers vote for Exclusion) (see grey colour cells in Table 6).

*Table 6   Dealing with disagreements*

|  |  | Reviewer X | | |
|---|---|---|---|---|
|  |  | Include | Uncertain | Exclude |
| Reviewer Y | Include | A | B | D |

|   | Uncertain | B | C | E |
|---|---|---|---|---|
|   | Exclude | D | E | F |

Table 7 shows the results of the selection process. The full references of the selected papers are listed in Appendix I.

*Table 7   Results of the selection process*

| Source | #works | Selected Works (PaperID) |
|---|---|---|
| Automated search | 29 | S1-S29 |
| Backward snowballing | 9 | S30-S38 |
| Forward snowballing | 10 | S39-S48 |

### 3.4. Data extraction

We developed a Data Extraction Form (DEF) to gather all the data extracted from the selected primary works. A template was used to organize the items of interest. Reviewers, who performed the data extraction process, added a comment to every cell with the rationale for their decision. The data extraction form, plus the search results and all the references, are offered online to interested readers as complementary material to this study (http://doi.org/10.6084/m9.figshare.7959425).

In addition to this, we devised a protocol to reduce individual bias (a frequent validity threat) during the data extraction process. The protocol stablished the following steps:

1. The set of selected primary works should be divided into two halves (H1 and H2),
2. Two reviewers (r1 and r2), independently, will extract data from H1,
3. A reviewer (r0) will compare outputs from reviewers, integrate results and resolve possible conflicts,
4. Step 2 is repeated with reviewers r1 and r2 working in H2. Finally, step 3 will be executed again until an agreement is reached.

We have clustered the data extracted to answer RQ1 into eight categories. Table 8 shows the primary selected studies belonging to each of these categories. The meaning of the identified categories is as follows:

- Estimation: the usefulness of Use Cases for assessing the early project estimate of the cost and effort;
- Analysis: better communication between stakeholders thanks to the structure and semantics of Use Cases. Consistency improvement (semantic and syntactic) among different models and supports in gathering a complete set of requirements for the system under construction;
- Automation: increase the level of automation thanks to the textual content described in the Use Case specification (e.g., test case generation, OCL constraints, domain analysis models, analysis of requirements etc.);
- Testing: facilitates the derivation of test information (executable test cases, test scenarios, test inputs) for different levels of testing;
- Reuse/SPL: the functionalities specified in the Use Cases can be reused in different applications. Some of them can be useful to develop new software product lines or improve existent ones;
- Quality: improvements derived from the assessment of Use Cases, with the purpose to provide defect data, identification of missing requirements and their analysis, conflict detection and so on;

- Traceability: the structure and semantics of Use Cases, as well as the relationships between different Use Cases, can improve traceability among derived software artefacts;
- Review/Validation: The structure and semantics of Use Cases help in reviewing/validating the functional requirements of the software system, from different perspectives, by applying various quality measures.

## 4. Results and discussion

Results from the data extraction process are reported and analysed in this section. We found a duplicate paper in the original set of selected papers (48 papers) while conducting the data extraction, so our final set consisted of 47 primary works (they are listed in Appendix I).

The systematic mapping study was conducted following the guidelines in [19,23]. For reproducibility of this study, care was taken to document the steps of the research protocol. Two complementary search strategies were used: a) automated search conducted in four publishers and indexers databases and b) backward and forward snowballing. The search strategies retrieved 5,805 papers and a set of 47 primary works were selected after the application of the inclusion/exclusion criteria. This relatively low number of selected works raises questions as to whether the results can be objectively generalized (a validity threat). Fortunately, the data extraction showed that the sample of selected papers had a wide coverage, including a good representation of advantages, disadvantages, application domains (industries), years, authors and organizations.

The results presented in the following sections may seem superficial or obvious. However, the goal of an SMS is to collect evidence that can corroborate or disprove beliefs about reality. Under this premise, we have reported the evidences found, objectively, following a systematic method. The following results are based on evidence, not false beliefs or personal opinions.

### 4.1. Which are the most frequently reported advantages/strengths of using Use Cases in software development in real-world settings?

We classified the reported benefits in eight categories, regarding the activity where the application of Use Cases have had the impact (see Table 8).

*Table 8   Categories of identified benefits*

| Advantages | Paper ID | #papers |
|---|---|---|
| Estimation | SP3, SP4, SP5, SP6, SP10, SP12, SP15, SP24, SP25, SP26, SP28, SP29, SP30, SP32, SP36, SP37, SP43, SP45 | 18 |
| Analysis | SP1, SP2, SP8, SP9, SP18, SP27, SP31, SP34, SP35, SP39, SP41, SP42, SP46 | 13 |
| Automation | SP17, SP19, SP20, SP22, SP35, SP40, SP45, SP46 | 8 |
| Testing | SP17, SP18, SP21, SP24, SP28, SP33, SP35 | 7 |
| Reuse/SPL | SP13, SP22, SP38, SP41, SP46 | 5 |
| Quality | SP16, SP23, SP44, SP47 | 4 |
| Traceability | SP33, SP39, SP47 | 3 |
| Review/validation | SP40, SP44, SP47 | 3 |

Most articles report a positive impact of using Use Cases for Estimation. The Use Case Points (UCP) method is the most cited for effort estimation [SP3, SP6 and SP29], for effort estimation to complete tasks /

workload [SP5 and SP10] or to estimate cost [SP12] or size [SP45]. Some authors have adapted the UCP technique for estimating effort [SP25], costs [SP30] or improving the estimation process [SP32]. The results also point to the advantages of Use Cases in other estimation methods, such as Weighted Nageswaran method for estimating test activity effort [SP4], a new method to estimate project size [SP15 and SP43], testing size [SP24], test execution effort [SP28], development effort [SP36 and SP26] and maintenance effort [SP37]. In the most recent study [SP5], the authors reported about the correlation between project sizes and effort required to complete the tasks.

The second category with the most benefits reported was Analysis. Authors reported aspects such as better understanding of customer needs [SP1, SP8, SP34 and SP39], including business requirements [SP2 and SP27] and compliance standards [SP46], as well as better understanding of the system [SP9]. The advantages of Use Cases include the support to ensure consistency between models [SP31, SP41 and SP42] and the management of uncertainties [SP18].

The benefits of Use Cases for automation are related to the generation of tests [SP17, SP19 and SP35], OCL constraints [SP20], generation of product-specific Use Cases in the context of SPL [SP22], generation of behaviour specifications [SP40], effort estimation [SP45] and security threats [SP46].

Seven studies report benefits of applying Use Cases to software testing in real-world applications. In [SP17] the authors propose to create a scenario template, defining possible user actions for each Use Case in Gherkin format, and in this way, enabling the automated generation and execution of test cases that reproduce failures. Use Cases also serve as a reference to generate test models [SP18], explicitly specifying uncertainty. In [SP35], the authors demonstrate that Use Cases can be employed in Restricted Natural Language Text; facilitating the derivation of test cases. Use Case specifications can be used to identify the test inputs that fire the state transitions in timed automata [SP21]. Effectiveness of Use Cases in test effort estimation is explored by [SP28]. Other benefits can be derived from the approach on Use Case verification points model [SP24] by ensuring traceability between requirements and system test cases; Use Cases can provide behavioural information that helps in deriving quality test cases [SP33].

The areas of Reuse and Software Product Lines (SPL) also obtain advantages when adopting Use Cases [48]. Use Cases are able to describe the common characteristics of all the products belonging to a product line and the variations that differentiate them [SP13 and SP38]. In [SP22] the authors propose a Use Case-centric product line modelling method that automated generation of product-specific Use Cases and domain models. The work in [SP41] describe a well-defined procedure for deriving functional architecture from Use Cases that can lead to reusable models. Finally, in [SP46], authors integrate an existing approach for modelling security and privacy requirements, in terms of security threats, their mitigations, and their relations to Use Cases in a misUse Case diagram. The key characteristic of this method is that it captures threat scenarios and mitigation schemes in an explicit and structured form, thus enabling both automated analysis of threat scenarios and reuse of mitigation schemes.

The results also point to gains related to traceability. In [SP33], authors highlight that Use Cases ensure traceability between requirements and system test cases, and, when tests are generated, it is possible to generate traces back to associated requirements [SP47]. In [SP39] the use of Use Case model is extended to include ranking information associated with each Use Case, and in this way, provides strong traceability and improvement in the understandings of developers and customers.

Some articles indicate that Use Cases facilitate the use of formal verification techniques [SP40], improve the process of problems/errors identification [SP44] and facilitates the review process [SP47].

Finally, we have found some advantages related to quality improvement. The effectiveness of Use Cases in improving the comprehension and the quality of the produced requirements specification is reported by [SP16]. The formalization of UCs as Graph Transformation models (GTs) help improve the quality of the Use Cases themselves [SP44]. In [SP47], the authors describe experiences for creating system test cases using Use Case models. They argue for the application of Use cases in early steps of requirements engineering to ensure the testability of requirements and, by doing that, improve the overall quality of general customer Use Cases. Paper [SP23] presents a quality assessment of Use Cases with the aim to provide previously identified defect data, allowing other industries to avoid the same (or similar) defects.

Three papers did not highlight any advantage [SP7, SP11 and SP14], however, they pointed out some disadvantages (discussed below, in section 4.2).

**How have they evolved over time?**

The selected studies and the reported benefits are shown in Table 9, sorted by year. The oldest study was published in 1997 [SP39]. It reports benefits in the categories of analysis and traceability. Since then, we have identified a continuous flow of studies. The year 2008 stands out with the highest quantity of publications reporting advantages, specially the positive effects in estimation (5 studies).

Table 9    Distribution of papers mentioning benefits along the years

| Benefit / Year | Automation | Analysis | Estimation | Quality | Reuse / SPL | Review / Validation | Testing | Traceability |
|---|---|---|---|---|---|---|---|---|
| 1997 | | SP39 | | | | | | SP39 |
| 1999 | | SP27, SP34 | | | | | | |
| 2001 | | | SP29 | | | | | |
| 2002 | | SP9 | SP32 | | | | | |
| 2003 | | SP1 | | | SP13 | | | |
| 2004 | SP45 | | SP45 | | | | | |
| 2005 | | | SP3, SP12, SP25 | | | | | |
| 2006 | | | SP43 | SP23 | | | | |
| 2007 | | SP42 | | | | | | |
| 2008 | | | SP6, SP24, SP26, SP28, SP36 | SP47 | | SP47 | SP24, SP28 | SP47 |
| 2009 | | | SP4, SP30 | | | | | |
| 2010 | | SP2, SP31 | SP15 | | | | | |
| 2011 | SP40 | | SP10, SP37 | | | SP40 | | |
| 2014 | SP35 | SP35, SP41 | | | SP41 | | SP35 | |
| 2015 | | SP8 | SP5 | SP44 | | SP44 | SP33 | SP33 |
| 2016 | SP19, SP22 | | | | SP22 | | | |
| 2017 | SP17 | | | | SP38 | | SP17, SP21 | |
| 2018 | SP20, SP46 | SP18, SP46 | | SP16 | SP46 | | SP18 | |

## 4.2. Which are the most frequently reported disadvantages/drawbacks of using Use Cases in software development in real-world settings?

Although Use Cases have been widely accepted and effectively used in many software development activities, several drawbacks have been also reported. Eighteen studies, among the selected 47, have reported disadvantages of using Use Cases in different activities of the software development process. Table 10 shows the categories, reference and quantity of studies reporting any disadvantages.

The category of Analysis contains the largest number of studies that report any disadvantage. The study SP7 reported an actor based strategy for identifying and defining Use Cases. However, not all Use Cases for all systems interact with external actors; there are systems that have significant functionality that is not a reaction to an external actor. Embedded control software systems provide a good example for such systems where major control functions are performed without significant external input. This makes the traditional Use Case technique less appropriate for such kinds of systems. They have also reported that the development of Use Cases require a detailed step-by-step guide to create a quality specification. The studies, SP8 and SP9 highlighted issues with the consistency of the terms, and level of detail of the Use Cases. The lack of guidance on how to analyse the roles, i.e. actors, of a subordinate system with respect to another are prone to errors [SP18, SP34 and SP44]. The adoption of Use Cases implies some costs (mainly training costs), because each iteration adds documentation, verification, and test costs [SP39, SP42].

Six studies reported issues related to Estimation. The studies highlighted the fact that the format and quality of the Use Case model have a direct impact in estimating the development efforts [SP11 and SP29]. Three studies point to the fact that the granularity of the actions [SP14, SP43 and SP45] described in the Use Cases steps were not uniform (some were too abstract and others were too detailed). One study [SP10] highlighted that the unavailability of a standardized format affects the estimation process and suggested that Use Cases have to be characterized in a consistent way.

*Table 10 Categories of reported disadvantages*

| Categories | Paper ID | #papers | Disadvantages |
|---|---|---|---|
| Analysis | SP7, SP8, SP9, SP18, SP34, SP39, SP42, SP44 | 8 | - The granularity of the actions described in the Use Cases steps was not uniform (some are too abstract and others too detailed).<br>- The difficulty of describing complete superordinate Use Cases for large complex systems (lack of guidance for analysing them).<br>- Cost of adoption and iterations. |
| Estimation | SP10, SP11, SP14, SP29, SP43, SP45 | 6 | - The format and quality of the Use Cases have a negative impact on the estimation process. This is due to the unavailability of standardized or step-by-step procedure/guidelines on writing the Use Cases |
| Testing | SP33, SP47 | 2 | - The flexibility and varied level of details among the Use Cases (i.e., the subordinates and super ordinates) make testing difficult.<br>- Adoption cost |
| Reuse/SPL | SP19, SP22 | 2 | - The effort of modelling the Use Cases to allow for reusability in different contexts.<br>- The variety in descriptions (level of details) |

In the testing category, two papers reported that the application of Use Cases require restrictive constraints to ensure the quality of the test cases [SP33], and have a high cost of adoption and iteration [SP47]. Concerning the Reuse/SPL category, two papers [SP19 and SP22] highlighted that the reusability of the Use Cases is low, require a significant amount of modelling efforts, and suffer from the absence of a methodology that explicitly manages the uncertainty of the modelling of Use Cases at different levels.

**How have they evolved over time?**

The first reported disadvantage (1997) is related to the unavailability of a standard format or step-by-step procedure for authoring Use Cases. In other words, the varied level of details among the Use Cases affects its quality and the process of analysis. The reporting of this disadvantage continued over the years (i.e., 1999, 2002, 2007, 2015 and 2018). Another disadvantage frequently mentioned, is the granularity when describing Use Case scenarios. This disadvantage of Use Case was reported over several years (2002, 2008 and 2015), and continued due to the unavailability of an automated Use Case analysis approach. The lack of quality assessment criteria and analysis process/guidelines was another disadvantage reported in 2007, 2002 and 2015. The cost of iterating Use Cases was first reported in 1997, together with their adoption cost. Recently, in 2016, the studies had also highlighted that the modelling effort of Use Cases is high and time-consuming. Table 11 shows the evolution of the number of papers reporting all these disadvantages over the years, clustering by category.

*Table 11  Distribution of papers mentioning disadvantages along the years*

| Benefit / Year | Analysis | Estimation | Testing | Reuse / SPL |
|---|---|---|---|---|
| 1997 | SP39 | | | |
| 1999 | SP34 | | | |
| 2001 | | SP29 | | |
| 2002 | SP7, SP9 | SP11 | | |
| 2004 | | SP45 | | |
| 2006 | | SP43 | | |
| 2007 | SP42 | | | |
| 2008 | | SP14 | SP47 | |
| 2011 | | SP10 | | |
| 2015 | SP8, SP44 | | SP33 | |
| 2016 | | | | SP19, SP22 |
| 2018 | SP18 | | | |

### 4.3. Which are the industry's domains?

Our study identified a large sample of domains, grouped into seven main categories (Table 12). The number of projects in the Information Technology category (Code 45) stands out with 20 projects. This category includes the areas of Software and Services (Code 4510) and Hardware and Equipment Technology (Code 4520). Two other categories that stand out are: 1) Consumer discretionary, which includes seven projects in the areas of Automobiles & Components (2510) and Retailing (Internet applications, 255020); and 2) Financials, including 4 projects in the area of Banks (4010) and 5 in the area of Diversified Financials (4020).

*Table 12  Industrial domains reported*

| GICS Category and Code | PaperID |
|---|---|
| Energy (10) | SP19, SP35 |
| Industrial (20) | SP6, SP7, SP9, SP18 |
| Consumer discretionary (25) | SP11, SP20, SP21, SP22, SP23, SP33, SP38 |
| Health care (35) | SP18, SP31, SP46 |
| Financials (40) | SP1, SP2, SP4, SP27, SP29(3 projects), SP30, SP42 |
| Information Technology (45) | SP3, SP10, SP13, SP15, SP16, SP19(2 projects), SP24, SP26(2 projects), SP28, SP32, SP35, SP36, SP37, SP39, SP40, SP44, SP45, SP47 |
| Telecommunication (50) | SP17, SP25, SP34 |

Figure 2 shows a summary of the identified categories and the quantity of mentions for each one. The first two digits of the categories in Figure 2 reflects the seven main domains reported in Table 12. The list with the full names of the categories in the GICS standard can be accessed online from [36].

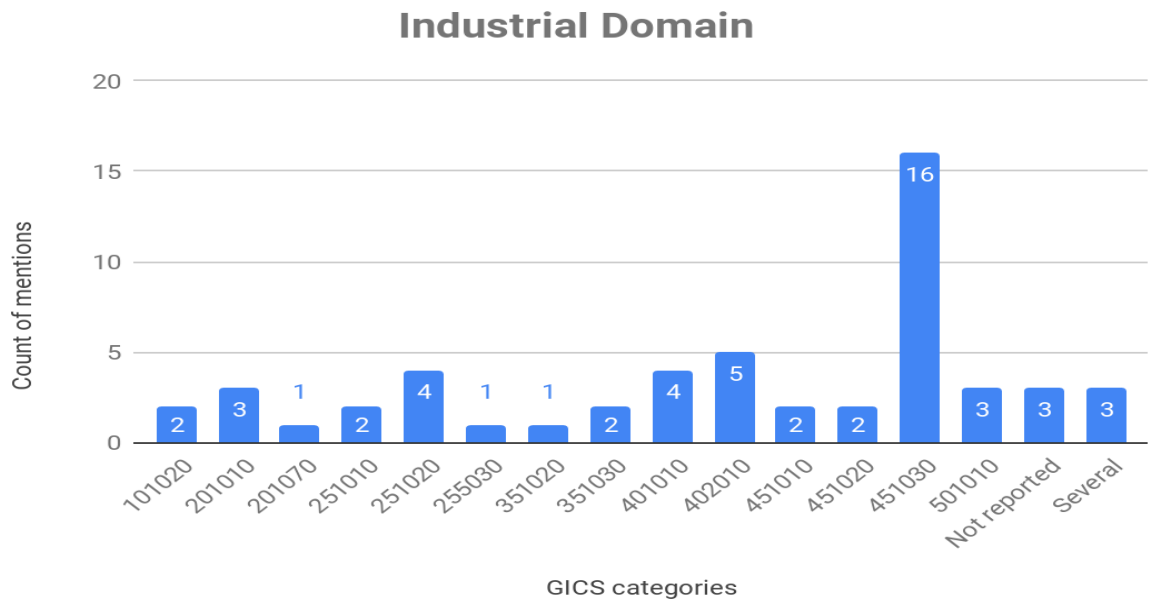

*Figure 2   Domain categories*

Three papers include the analysis of several projects [SP5, SP12 and SP14], related to various industrial domains. In those cases we have reported the domain in the "Several" category, to the right of the graph. Three other papers did not report any data about the domain [SP8, SP41 and SP43].

### 4.4. Which research types were applied?

We used the classification proposed by Wieringa [39] and have followed the guidelines in [23] to classify the research type reported in the selected papers.

As we were interested in papers reporting real-world experiences, only three research types were considered: evaluation research, experience report and solution proposal. Figure 3 shows the quantity of papers reporting each research type, while Table 13 provides their references.

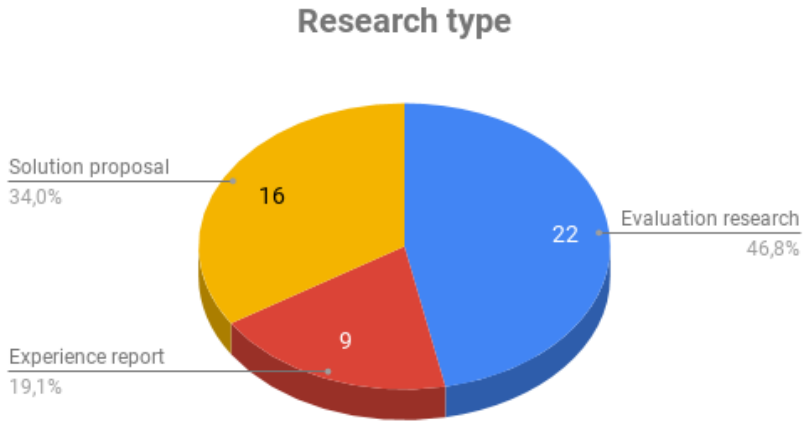

*Figure 3  Distribution of papers by research type*

Evaluation research includes two main activities, the problem investigation (hypothesis) and the implementation evaluation (experimentation). Practitioners, describing a personal experience using a specific technique or tool, usually write Experience reports. Finally, Solution proposals argues for the relevance or advantages of a novel technique or tool. Table 2, in the Introduction section, shows the conditions that papers reporting these types of research should fulfil.

*Table 13  Research type of selected papers*

| Experience report | Evaluation research | Solution proposal |
|---|---|---|
| SP1, SP6, SP7, SP9, SP10, SP24, SP29, SP34, SP41. | SP3, SP4, SP5, SP11, SP16, SP18, SP19, SP21, SP23, SP25, SP28, SP32, SP37, SP38, SP39, SP40, SP42, SP43, SP44, SP45, SP46, SP47. | SP2, SP8, SP12, SP13, SP14, SP15, SP17, SP20, SP22, SP26, SP27, SP30, SP31, SP33, SP35, SP36. |

It is important to note that the Evaluation Research type requires an empirical evaluation. This evaluation can be done by any of the following methods:

- Industrial Case study,
- Controlled experiment with practitioners,
- Practitioner targeted survey,
- Action research, or
- Ethnography.

In general, the quality of the selected papers in this study is high, given the type of research reported. The fact that almost half of the papers (46.8%) applied Evaluation research is a clear symptom of the strong scientific basis, and therefore, of the reliability of the results. On the other hand, the presence of an important percentage of papers reporting Solution proposals (34%), indicate that there are still gaps in the research, and that the application of Use Cases to projects in real world projects is not yet a mature area.

An annoying outcome of this study was the lack of raw data reports. None of the selected papers offered access to this type of information, so the possibility to replicate their studies was drastically reduced.

## 4.5. Were threats to validity reported?

We have used the classification of validity threats in software engineering proposed by Petersen and Gencel [49], based on the Pragmatic Software Engineering View.

Only 15 (one third) out of the 47 papers analysed, reported validity threats. Figure 4 shows the distribution of the validity threats (categories) reported in the primary studies.

It is important to note that, with the exception of the descriptive category; the other categories are evenly distributed among the selected papers. The report of the descriptive validity, which implies the objective description -without bias- of the characteristics, and especially of the conclusions, is not frequent in the set of selected papers (11.8%). The authors of these papers assume that their interpretations are correct and accurate, and rarely offer verification mechanisms.

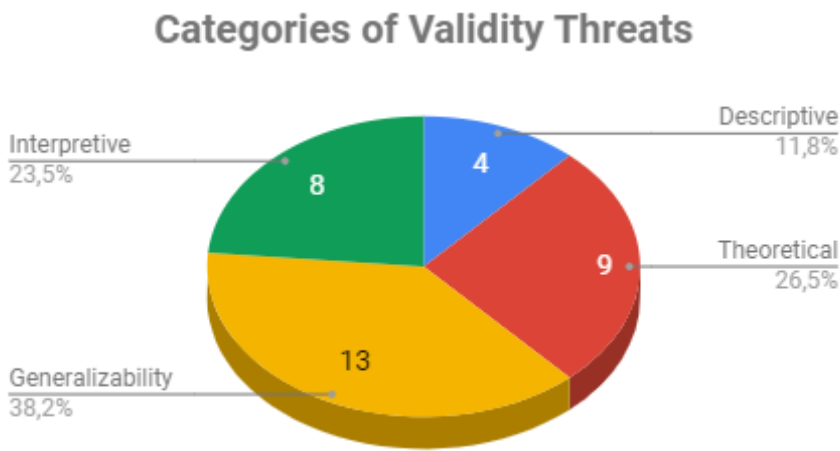

*Figure 4   Types of validity threats reported*

Many authors mention the risk involved in assuming the generalization of their results (38.2%), and are concerned with highlighting those uncontrollable factors that could affect the experience and the results. Finally, we have identified that in eight papers (23.5%) the authors report verification mechanisms to reduce the bias in the derivation of the conclusions, and try to demonstrate the logical relationship between the data and the inferences they have made.

In summary, we think that the quality of the reporting should be improved. The addition of a detailed report about the threats to validity, and how they were eliminated (or their impact was reduced), would improve the overall quality of the paper and facilitates the understanding and interpretation of the data. We believe that the development of a generic template that helps researchers to fill out this kind of reports could be of great help. On the other hand, chief editors of publishing journals can influence on the authors, by encouraging them to include a report of validity threats in their papers, as a prerequisite for publication (as well as a structured Abstract).

## 4.6. Demographic questions

The following subsections present the results of our study regarding the demographic aspects, such as active researchers, organizations and countries, as well as the top publishing venues.

**DQ1: Who are the most active researchers?**

Based on the analysis of 47 primary studies, we found that 124 researchers have contributed on the Use Case research. For those researchers authoring more than one paper, Table 14 shows the researchers' name and the number of papers in which they have participated.

*Table 14  Active researchers*

| Researchers | Count |
|---|---|
| Briand, L.C. | 6 |
| Goknil, A.; Pastore, F. | 4 |
| Ali, S.; Bente, A.; Wang, C.; Yue, T.; Zhu, X. | 3 |
| Bernat, G.; Chen, L.; Hajri, I.; Leotta, M.; Liu, C.; McDermid, J.; Nasr, E.;  Qu, Y.; Reggio, G.; Ricca, F.; Thierry, S.; Wang, F.; Zhang, H.; Zhang, M.; Zhou, B. | 2 |

**DQ2: Which are the most active organizations?**

We extracted the affiliation of each author from all the 47 selected studies. As we can see in Figure 5, the largest number of authors (68.9%) are from the academy arena, with few authors affiliated with both, academia and industry (only 6 entries).

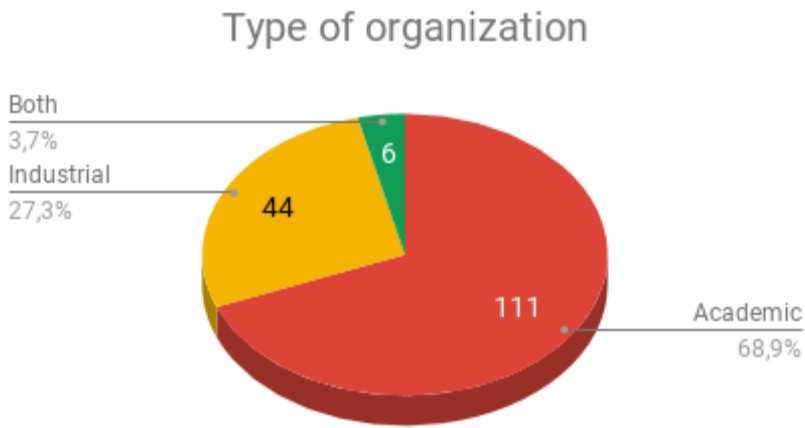

*Figure 5  Distribution of the type of organization*

Table 15 list the top five academic institutions with the largest presence among the authors' affiliations and the top four in the industry context. The University of Luxembourg is the institution that appeared more frequently among the authors' affiliations, 21 in total, followed by Zhejiang University, with 10. Simula Research Laboratory leads the list among the affiliations considered as industrial, with 13 affiliations.

*Table 15  Most active organizations (Academic & Industrial)*

| Top five Academic | Count | Top four Industrial | Count |
|---|---|---|---|
| University of Luxembourg | 21 | Simula Research Laboratory | 13 |
| Zhejiang University | 10 | Hitachi Software Engineering Co., Ltd. | 5 |

| University of Oslo | 8 | Siemens AG | 4 |
| --- | --- | --- | --- |
| York University | 6 | IBM China Research Lab | 3 |
| Federal University of Rio Grande do Sul | 6 | | |

**DQ3: Which are the most active countries?**

Table 16 shows the most active countries, related to authors' affiliations. Most of the papers come from authors in European countries (96 entries), followed by the Asian continent (41 entries).

China, Luxembourg and Norway are the countries that stand out most in this field. Some authors in China and Norway also present collaborative studies (SP19 and SP35). Researchers in Norway report results in the industry since 2001 (SP29), the Chinese in 2008 (SP24 and SP28) and their first collaborative study date from 2014 (SP35). The results of researchers from Luxembourg are more recent, starting in 2015 (SP33), but continuing to publish in 2017 (SP21 and SP38) and 2018 (SP20, SP22 and SP46).

*Table 16  Top eight active countries*

| Country | China | Luxembourg | Norway | Brazil | Germany | UK | Japan | Italy |
| --- | --- | --- | --- | --- | --- | --- | --- | --- |
| Count | 28 | 23 | 20 | 16 | 11 | 10 | 10 | 10 |

**DQ4: Which are the Top publication venues?**

Thirty-seven studies were published in conferences and workshops, and the remaining 10 studies were published in journals.

The findings revealed that there is no a single venue that can be considered as the top one to publish research related to Use Cases. Table 17 shows the top conference publication venues (the remaining venues have only a single count). Regarding to the Journal publication venue, 10 studies were published in 9 different Journals and among them only the "Journal of Systems and Software" had 2 entries, while the others have only a single count.

*Table 17  Top Conferences/Workshops of published reports*

| Conference/workshop | Count |
| --- | --- |
| International Conference on Software Testing Verification and Validation | 4 |
| International Conference on Product Focused Software Process Improvement | 2 |
| International Symposium on Empirical Software Engineering | 2 |
| International Workshop on Object-Oriented Real-Time Dependable Systems | 2 |

## 4.7. Cross analysis

Two aspects stand out when we analysed the domain of industry (RQ3) in which the reported advantages (RQ1) were identified. First, the largest number of studies reports advantages in the area of estimation (10 papers), while most of the papers report experiences in industries related to the domain of Information Technology (IT). On the other hand, this domain of IT is the only one that includes advantages reported in all the categories studied (Figure 6, penultimate column).

The financial domain (first column) was the one that presented benefits related to fewer categories (only 2, Analysis and Estimation), however, seven studies (14.9%) reported experiences in this domain. The Health care domain (right side of the graph) presented benefits in three categories, but including only two studies (SP31 and SP46).

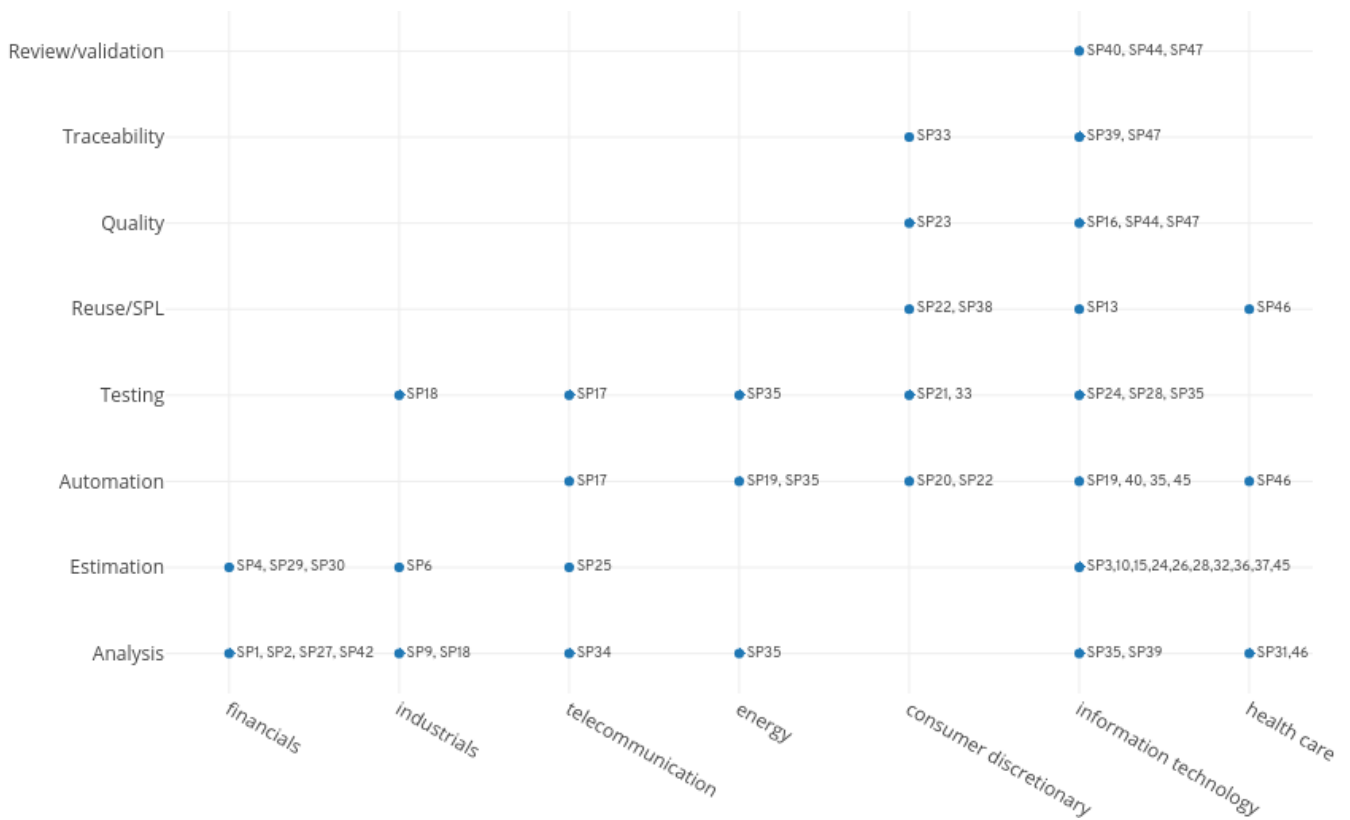

*Figure 6 Industrial domain and reported advantages*

We have analyzed the disadvantages (RQ1) against the domain of industry (RQ3) and found that the largest number of papers reporting drawbacks occurred in the software domain (6 studies, 12.8%) under all of the four categories: analysis, estimation, testing and reuse/SPL. The correlation between the industrial domains and disadvantages reported by the selected studies is shown in Figure 7. The figure shows that most of the papers reported disadvantages in the analysis category, the disadvantages have been reported in six different domains (exceptions were Textile and Automobile). The less number of papers belongs to the categories of Testing and Reuse/SPL (two papers each, in the Software and Automobile domains). Three domains share the least number of papers reporting disadvantages: Health care (SP18) and Telecommunications (SP34), both in the Analysis category, and Textile (SP11) in the Estimation category.

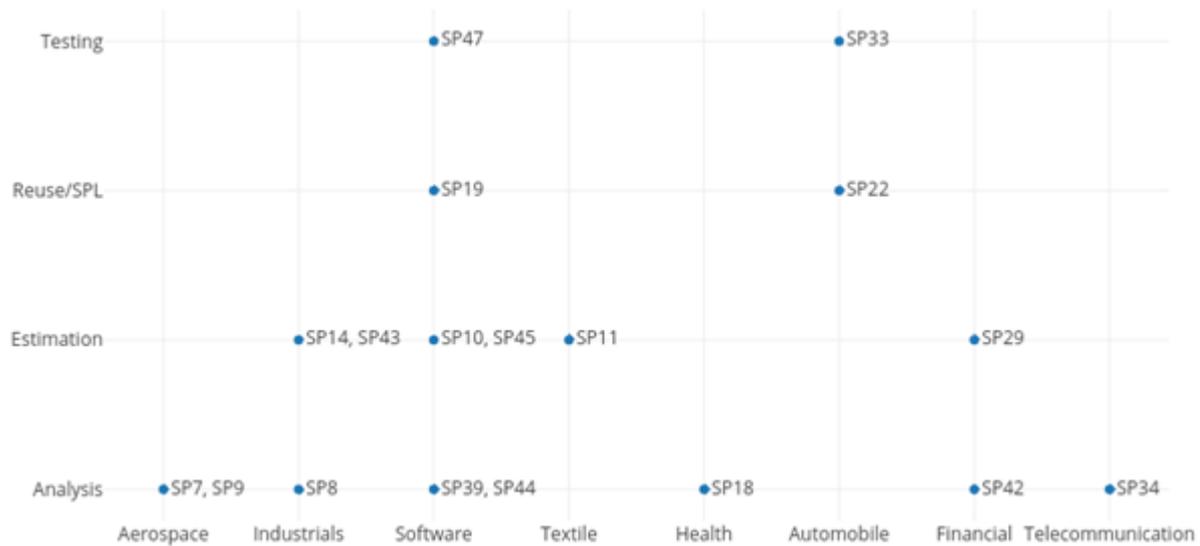

*Figure 7     Industrial domain and reported disadvantages*

## 5. Validity threats

We considered the four types of possible threats to validity mentioned in [49] during the design of the SMS protocol, and mitigated them as follows:

- Descriptive validity: *"the extent to which observations are described accurately and objectively"* [49]. To reduce this threat, we designed a DEF that objectively quantifies the data extraction process. By allowing for public access to this DEF, external reviewers can access it and revise the original data. Therefore, we considered this threat to be controlled. The DEF is available online as complementary material.
- Theoretical validity: *"the ability of being able to capture what we intend to capture"* [49]. Here we considered two activities:
  o Study identification/sampling: (missed studies). To reduce this threat, we applied four search strategies: 1) automated search, 2) backward and forward snowballing, 3) manual search of publications of the most cited authors and, 4) a manual search of works published in well-known international conferences. As each search strategy could introduces bias due to the source of data, we argue that this approach leads us to a fair sample of complementary sources.
  o Data extraction and classification: we apply the protocol detailed in Section 3.
- Interpretive validity: *"achieved when the conclusions drawn are reasonable given the data"* [49]. We applied the same strategy as that used for data extraction, dividing the work in two halves and having pairs of reviewers performing overlapping analysis. To deal with disagreements, we applied the criteria described in Table 6.
- Repeatability: *"requires detailed reporting of the research process"* [49]. We offered a detailed report of the research process and all the data gathered during our study. The detailed protocol in section 3, and the data extraction form, available online for public access, facilitates the repeatability of this study.

In addition, the relatively low number of selected works raise questions about whether the results could be generalized. In particular, after having reviewed our process, we believe that, despite the relatively large body of knowledge on Use Cases, the reported results based on solid evidence are by far scarce. This fact is enhanced by the focus of this research (limited to their application in industrial contexts).

Finally, we invited two external researchers to review and validate our research protocol (section 3) in order to obtain feedback and opportunities for improvement. The output of this external review was a template, adapted from [23], to conduct a self-assessment of the work done to "systematically develop" this SMS. The template can be used as a self-evaluation tool, to help authors check whether their methodology has been done correctly (see Appendix II).

# 6. Conclusions and future work

This paper has presented a systematic mapping study on the application of Use Cases in real world projects. Our work identified reported advantages of using Use Cases in 8 categories and 7 industrial domains. The category with the highest number of reported advantages was *Estimation*, and the most cited industry domain was I*nformation Technology*. The earliest study identifying advantages due to the application of Use Cases was published in 1997, and since then, we can observe a continuous flow of studies reporting positive outcomes.

On the other hand, the reported drawbacks of using Use Cases were classified in four different categories. Specifically, the larger number of disadvantages were reported under the *Analysis* and *Estimation* categories. The two most mentioned disadvantages were the granularity of describing scenarios and lack of analysis guidelines/processes. Many authors reported issues such as inconsistency, incorrectness, and incompleteness, associated to the use of natural language when specify Use Cases. Several efforts have been made to formalize (or restrict) the specification of requirements by means of Use Cases documentation, but the lack of a standard format and their heterogeneous granularity is a concern that still needs to be addressed. Additionally, the availability of quality assessment criteria and the automated analysis of Use Cases would make them more reusable, maintainable and applicable in different stages of the development process.

Twenty-two works (47%) applied evaluation research as defined in [39]. This type of research implies use in practice and empirical evaluation, which implies a high level of quality in the reported research. Nine works (19%) were experience reports, in which the evaluation mechanism is not always reported. We believe that these percentages reflect a good level of scientific quality in the set of papers selected for this study. More research is needed to enforce the application of higher quality research approaches (guidelines, examples, data repositories, reporting templates and so on).

Although it is nice to highlight the positive outcomes of a technique or a solution proposal, we believe that it is more important, for researchers and professionals, to be warned of the negatives. Our study can serve as a guide for professionals to make better decisions, knowing in advance the possible negative effects of the technique they plan to apply. On the other hand, researchers can find inspiration to initiate studies aimed at solving current problems.

Some open issues for future research include:

- To conduct a systematic literature review to deepen the knowledge about the processes where Use Cases have been applied and how the benefits were measured (metrics).
- The number of selected works for this study is small. This provides an opportunity to investigate the

development of simpler protocols that can be used in industry to gather relevant data in an easier way, while applying rigorous methods.
- Develop and validate a model suitable to link Use Case advantages to economic values (strategic or financial).
- We found that the standard taxonomy for industrial domains is not well suited for software engineering. It was difficult to classify some applications, and many of the categories were never used. An interesting future work can be the development of a taxonomy of industrial domains more appropriate for software engineering and software development projects.
- Another interesting proposal for future work is to develop a "standard" template to report validity threats (according to the research method applied), to facilitate this task to researchers and increase the scientific rigor of their reports.

**Acknowledgments**

We are grateful for the constructive feedback from the anonymous reviewers of this work who helped improving it significantly.

**Appendices**

### A. Selected (included) studies

| PaperID | Full reference |
|---|---|
| SP1 | Patrício L., Cunha J.F., Fisk R.P., Pastor O. (2003) Essential Use Cases in the Design of Multi-channel Service Offerings — A Study of Internet Banking. ICWE 2003. Lecture Notes in Computer Science, vol 2722. Springer, Berlin, Heidelberg |
| SP2 | D. Stock, F. Wortmann, J.H. Mayer, Use Cases for business metadata--a viewpoint-based approach to structuring and prioritizing business needs, in: Int. Conf. Bus. Inf. Syst., 2010: pp. 226–237. |
| SP3 | B. Anda, H.C. Benestad, S.E. Hove, A multiple-case study of software effort estimation based on Use Case points, in: Empir. Softw. Eng. 2005. 2005 Int. Symp., 2005. |
| SP4 | É.R.C. de Almeida, B.T. de Abreu, R. Moraes, An alternative approach to test effort estimation based on Use Cases, in: Softw. Test. Verif. Validation, 2009. ICST'09. Int. Conf., 2009: pp. 279–288. |
| SP5 | J. Popovic, D. Bojic, N. Korolija, Analysis of task effort estimation accuracy based on Use Case point size, IET Softw. 9 (2015) 166–173. |
| SP6 | C.M.B. da Silva, D.S. Loubach, A.M. da Cunha, Applying the Use Case points effort estimation technique to avionics systems, in: Digit. Avion. Syst. Conf. 2008. DASC 2008. IEEE/AIAA 27th, 2008: p. 5--B. |
| SP7 | E. Nasr, J. McDermid, G. Bernat, Eliciting and specifying requirements with Use Cases for embedded systems, in: Object-Oriented Real-Time Dependable Syst. 2002.(WORDS 2002). Proc. Seventh Int. Work., 2002: pp. 350–357. |

### B. Mapping process evaluation

This section contains an evaluation of the work done to "systematically develop" this SMS. It can be used as a self-evaluation, to help authors to check if everything has been done in the right way. The next table summarizes all the possible activities to consider when conducting a systematic mapping study.

**Identified activities for conducting a Systematic Mapping Study. Adapted from** [23]

| Phase | | Actions | Applied |
|---|---|---|---|
| Need for map | | Motivate the need and relevance | √ |
| | | Define objectives and questions | √ |
| | | Consult with target audience to define questions | --- |
| Study Identification | | | |
| | Choosing search strategy | Automated search (databases) | √ |
| | | Snowballing | √ |
| | | Manual (Conferences, Main Authors) | √ |
| | Develop the search | PICO | √ |
| | | Consult librarians or experts | √ |
| | | Iteratively try finding more relevant papers | √ |
| | | Keywords from known papers | √ |
| | | Use standards, encyclopaedias, and thesaurus | --- |
| | Evaluate the search | Test-set of known papers | √ |
| | | Expert evaluates result | √ |
| | | Search web-pages of key authors | √ |
| | Inclusion/Exclusion | Identify objective criteria for decision | √ |
| | | Add additional reviewer, resolve disagreements | √ |
| | | Decision rules | √ |
| Data extraction and Classification | | Identify objective criteria for decision | √ |
| | | Obscuring information that could bias | --- |
| | | Add additional reviewer, resolve disagreements | √ |
| | | Test–retest | --- |
| | | Classification scheme | √ |
| | | Research type | √ |
| | | Research method | √ |
| | | Venue type | √ |
| Validity discussion | | Validity discussion/limitations provided | √ |

We applied the evaluation rubric suggested by Petersen [23] to evaluate our work in terms of all the key activities a SMS should include. The following tables show the rubric criteria. The scores identified by our mapping study are highlighted (bold text):

**Rubric: need for review.**

| Evaluation | Description | Score |
|---|---|---|
| No description | The study is not motivated and the goal is not stated | 0 |
| Partial evaluation | Motivations and questions are provided | 1 |
| Full evaluation | Motivations and questions are provided, and have been defined in correspondence with target audience | **2** |

**Rubric: choosing the search strategy.**

| Evaluation | Description | Score |
|---|---|---|
| No description | Only one type of search has been conducted | 0 |

| Minimal evaluation | Two search strategies have been used | **1** |
|---|---|---|
| Full evaluation | Three or more search strategies have been used | 2 |

**Rubric: evaluation of the search.**

| Evaluation | Description | Score |
|---|---|---|
| No description | No actions have been reported to improve the reliability of the search and inclusion/exclusion | 0 |
| Minimal evaluation | At least one action has been taken to improve the reliability of the search OR the reliability of the inclusion/exclusion | 1 |
| Partial evaluation | At least one action has been taken to improve the reliability of the search AND the inclusion/exclusion | 2 |
| Full evaluation | All actions identified have been taken | **3** |

**Rubric: extraction and classification.**

| Evaluation | Description | Score |
|---|---|---|
| No description | No actions have been reported to improve on the extraction process or enable comparability between studies through the use of existing classifications | 0 |
| Minimal evaluation | At least one action has been taken to increase the reliability of the extraction process | 1 |
| Partial evaluation | At least one action has been taken to increase the reliability of the extraction process, and research type and method have been classified. | **2** |
| Full evaluation | All actions identified have been taken | 3 |

**Rubric: study validity.**

| Evaluation | Description | Score |
|---|---|---|
| No description | No threats or limitations are described | 0 |
| Full evaluation | Threats and limitations are described | **1** |

Our SMS obtained a final score of 9/11